\documentstyle[epsfig]{aipproc}

\def\frac#1#2{ {{#1} \over {#2} }}

\def\abs#1{\left| \: #1 \: \right|}%
\def\bom#1{\mbox{\boldmath$#1$}}
\def\beq{\begin{displaymath}}
\def\eeq{\end{displaymath}}

\def\as{\alpha_S}
\def\asb{\bar \alpha_S}

\def\bkq{\abs{\bom{k}+\bom{q}}}
\def\om{\omega}
\def\ga{\gamma}
\def\tga{\tilde \gamma}
\def\tchi{\tilde \chi}

\def\De{\Delta}

\def\cA{{\cal A}}
\def\Q_s{\mu}

\def\np#1#2#3{Nucl.\ Phys.\ {\bf B#1}, #2 (19#3)}
\def\pl#1#2#3{Phys.\ Lett.\ {\bf #1B}, #2 (19#3)}
\def\pr#1#2#3{Phys.\ Rev.\ D {\bf #1}, #2 (19#3)}

\def\zp#1#2#3{Zeit.\ Phys.\ {\bf C#1}, #2 (19#3)}

\begin{document}
\title{Angular ordering and small-$\bom{x}$ structure functions.}

\author{Gavin P. Salam}
\address{INFN --- Sezione di Milano, Via Celoria 16, 20133 Milano, Italy}

\begin{flushright}
IFUM 565-FT \\
hep-ph/9705233
\end{flushright}

\maketitle

\begin{abstract}
This talks examines the effect of angular ordering on the small-$x$
evolution of the unintegrated gluon distribution, and discusses the
characteristic function for the CCFM equation.
\end{abstract}

For some time now it has been known that angular ordering
[1] is an essential element in any description of
small-$x$ final state properties [2]. As a first step of a
programme to study the final state in small-$x$ physics, one should
examine the effect of angular ordering on the small-$x$ evolution of
the gluon structure function. Phenomenological studies have already
been performed [3], but this talk will examine the solutions
of the CCFM equation [2] from a more theoretical point of
view.

\begin{figure}[t]\label{gpsfig:kinebw}
\begin{center}
\input{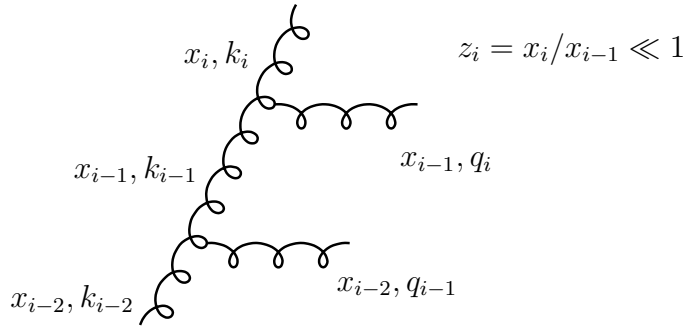}
\caption[]{Labelling of momenta}
\end{center}
\end{figure}

The main difference between the BFKL [4] and CCFM equations is
in the collinear region: in the BFKL case, the $i^{th}$ emission has
transverse momentum $q_i >\mu$, with $\mu\to0$; this regulates the
collinear divergence which is present, but gives the wrong final state
properties. In the CCFM case, angular ordering of emissions leads to
the following condition (see figure~1):

\begin{displaymath}\label{ao}
\theta_{i}>\theta_{i-1}\,,
\;\;\;\;\;\;\Rightarrow\;\;\;\;\;\;
q_{i} > z_{i-1} q_{i-1}
\,,
\end{displaymath}

\noindent with the corresponding gluon emission distribution being

\begin{displaymath}\label{dP}
d{\cal P}_i
=\frac{d^2q_i}{\pi q_i^2} \; dz_i\frac{\asb}{z_i}
\;\De(z_i,q_i,k_i)\;\Theta(q_i-z_{i-1}q_{i-1})
\,.
\end{displaymath}

\noindent The non-Sudakov form factor $\De$, which is analogous to a
probability for suppressing any further radiation, is defined by 

\begin{displaymath}\label{De}
\ln \De(z_i,q_i,k_i)=
-\int_{z_i}^1 dz' \;\frac{\asb}{z'}
\int\frac{dq'^2}{q'^2}\;\Theta(k_i-q')\;\Theta(q'-z'q_i)
\,.
\end{displaymath}

The elimination of a large fraction of the small-transverse-momentum
emissions means that angular ordering has a big effect on the final
state. But in structure function evolution, since collinear
singularities cancel, at leading order the BFKL and CCFM structure
functions are equivalent.

As part of a program to carry out a full investigation of the effects
of angular ordering at small $x$, this talk examines the component of
the next-to-leading order corrections to structure function evolution
that arise from angular ordering. Such effects are expected to be part
of the full NLO contribution [5].

Qualitatively since angular ordering reduces the phase space for
evolution, the exponent of the small-$x$ growth ought to be
reduced. The symmetry, present in the BFKL equation, between large and
small scales will be broken, favouring evolution to large momentum
scales. Finally diffusion will be reduced because large jumps (down)
in scale are suppressed.

There are two limits in which the effects of angular ordering should
disappear: as $\as \to 0$, because the typical $z_{i-1} \sim \as$ will
be very small (this justifies the assertion that for structure
functions the effects of angular ordering are next to leading); and in
the double-leading-logarithmic limit because the condition $q_i >
q_{i-1}$ automatically satisfies the angular ordering condition.

The analytic treatment of the CCFM equation is more complicated than
that of the BFKL equation because the gluon density contains one
extra parameter, $p$, which defines the maximum angle for the emitted
gluons. In DIS it enters through the angle of the quarks produced
in the boson-gluon fusion. The equation for the CCFM density,
$A(x,k,p)$, of gluons with longitudinal momentum fraction $x$ and
transverse momentum $k$ is:

\begin{displaymath}\label{A1}
\cA(x,k,p) \;=\; \cA^{(0)}(x,k,p) \;+\;
\int\frac{d^2q}{\pi q^2}\; \frac{dz}{z}
\;\frac{\asb}{z}\De(z,q,k)\;
\Theta(p-zq)\;\cA(x/z,k',q)
\,,
\end{displaymath}

\noindent where $k'=\bkq$. By analogy to the BFKL equation one can
develop some understanding of it by looking for eigensolutions
(strictly speaking eigensolutions of the equation without an
inhomogeneous term and with no upper limit in the $z$ integral) of the
form

\begin{displaymath}
x\cA(x,k,p) = 
x^{-\om} \frac{1}{k^2}\left(\frac{k^2}{k_0^2}\right)^\ga G(p/k)
\,,
\end{displaymath}

\noindent where $G(p/k)$ parameterises the unknown dependence on
$p$. For $0<\ga<1$, one obtains a coupled pair of equations for 
$G$ and $\om$:

\begin{displaymath}\label{dG}
p\;\partial_p\;G(p/k)=
\asb
\int_p\;\frac{d^2q} {\pi q^2}
\left(\frac{p}{q}\right)^{\om}\;\De(p/q,q,k)
\;G\left(\frac{q}{k'}\right)
\;\left(\frac{{k'}^2}{k^2}\right)^{\tga-1}
,
\end{displaymath}

\noindent with the initial condition $G(\infty)=1$ and

\begin{displaymath}\label{A5}
\om = \as\tchi(\tga,\as) = \as
\int\frac{d^2q}{\pi q^2}
\left\{
\left(\frac{{k'}^2}{k^2}\right)^{\tga-1}
  G\left(\frac{q}{k'}\right)
-\Theta(k-q)\; G(q/k)\right\}
.
\end{displaymath}

\noindent In the second of these equations, if $G=1$
one notes that $\tchi$ is just the BFKL characteristic function. Since
$1-G(p/k)$ is formally of order $\as$, this demonstrates that angular
ordering has a next-to-leading effect on structure function
evolution. One can also show that in the limit of $\ga \to 0$ the
difference $\chi(\ga) - \tchi(\ga,\as)$ tends to a constant, which
implies corrections to the small-$x$ anomalous dimension of the form
$\as^3/\om^2$.

\begin{figure}\label{gpsfig:cvg}
\begin{center}
\epsfig{file=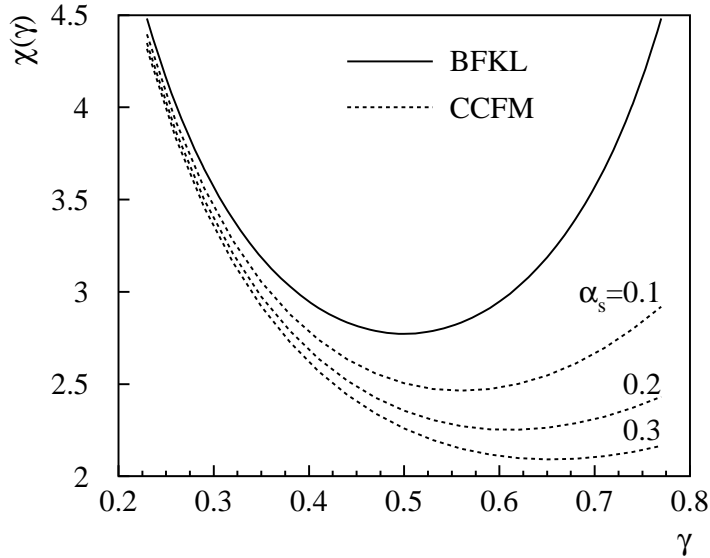, width=0.7\textwidth}
\caption[]{The BFKL and CCFM characteristic functions as a function of
$\ga$ for different values of $\as$.}
\end{center}
\end{figure}

Though a number of asymptotic properties of $G(p/k)$ have been
determined [6], it has not so far been possible to obtain its
full analytic form. Further understanding requires numerical
analysis. This has been carried out and figure~2 shows
the results for $\tchi$ compared to the BFKL characteristic function
for three different values of $\as$. It illustrates that as $\as\to0$
the two tend to coincide as happens also in the region $\ga \to0$ (the
DLLA region).

The loss of symmetry under $\ga\to1-\ga$ relates to the loss of
symmetry between small and large scales. Indeed, in contrast to the
BFKL case, there is no longer even a divergence at $\ga=1$.
Correspondingly, the minimum of the characteristic function gets
shifted to the right and is lower.

Particularly strong is the change in the second derivative, which for
$\as=0.2$ is reduced by a factor a two, indicating that for a given
amount of diffusion to occur, one needs twice the rapidity ($\ln x$
range) predicted by the BFKL equation, as seen for example in the
result of Mueller [7] for the $x$ value where the
operator-product expansion starts to break down due to diffusion,

\begin{displaymath}
\ln \frac{x_0}{x} \simeq \frac{1}{2 \tchi_c''}\ln \frac{Q^2}{\Lambda^2}
\,,
\end{displaymath}

\noindent where $\Lambda$ is the QCD scale, $x_0$ some starting point
for the small-$x$ evolution, and $Q^2$ the hard scale of the problem.

\section*{Acknowledgements}
This research was carried out in collaboration with G.~Bottazzi,
G.~Marchesini and M.~Scorletti and supported by funding from the
Italian INFN. We are very grateful to M. Ciafaloni, Yu.L. Dokshitzer,
A.H. Mueller and B.R. Webber for helpful discussions.

\end{document}